\documentclass[preprint,12pt]{elsarticle}

\usepackage{amssymb}
\usepackage{subfig}

\journal{Nuclear Instruments and Methods in Physics Research Section A}

\begin{document}

\begin{frontmatter}



\title{In-situ quantification of gamma-ray and beta-only emitting radionuclides}


\affiliation[inst1]{organization={Applied Nuclear Physics, Lawrence Berkeley National Laboratory},
            addressline={1 Cyclotron Rd}, 
            city={Berkeley},
            postcode={94720}, 
            state={California},
            country={USA}}
            
\affiliation[inst2]{organization={Department of Nuclear Engineering, University of California Berkeley},
            city={Berkeley},
            postcode={94720}, 
            state={California},
            country={USA}}

\author[inst1,inst2]{Kai Vetter}
\author[inst1]{Donald Gunter}
\author[inst1]{Paul Luke}
\author[inst1]{Victor Negut}
\author[inst1]{Ryan Pavlovsky}
\author[inst1]{Brian Plimley}
\author[inst1]{Joanna Szornel}

\begin{abstract}
The prompt and in-situ assessment of non-gamma ray emitting radionuclides such as Sr-90 remains an outstanding challenge, particularly in radiological emergency response and consequence management situations. We have developed a new concept to quantitatively assess a wide range of radionuclides, including beta-only emitters, using coplanar-grid CdZnTe detectors that provide depth-of-interaction sensing. By combining measurements with and without an electron absorber, we demonstrate the feasibility of detecting and identifying Sr-90 and other radionuclides with a sensitivity of about 1 $\mu$Ci/m$^2$ or 3.7x10$^4$ Bq/m$^2$, which is 10\% of the Derived Response Level, in less than 60 minutes. The new compact instrument can be used in the field or mobile laboratories to quickly assess a wide range of samples with sufficient sensitivity and specificity to provide critical guidance in the response after radiological incidents.
\end{abstract}


\end{frontmatter}


\section{Introduction}
\label{sec:introduction}
Current approaches for the quantification of pure beta-emitting radionuclides involve sending the sample to a remote laboratory for chemical separation and measurements, which can take up to two weeks. Alternatively, detection methodologies are available which are based on scintillators but are limited in specificity as many gamma ray-emitting radionuclides such as Cs-137 also decay via beta decay and are often mixed with beta-only emitters \cite{boja1990,iaea1995,vajda2010,bae2018,kang2020}. The assessment of radionuclides in general and of fission products such as Sr-90 specifically is important to determine and control the effective radiation dose to people and the environment. Of particular concern is the prompt assessment after the accidental or intentional releases of radionuclides to effectively guide emergency response actions. Gamma-ray detectors are readily available on local, regional, and national levels in order to detect illicit materials and to prevent their (mis)use or to respond to radiological incidents in-situ or in so-called Fly-Away Laboratory (FLA) and provide the necessary information typically within one hour after setup \cite{frmac2019}. Gamma-ray energies provide a powerful signature and fingerprint of specific radionuclides that enables the accurate assessment of material being detected or dispersed and allow the dose estimation and actions to minimize the radiation exposure. Instruments that enable the in-situ assessment of non-gamma ray emitting radionuclides with sufficient specificity are not available today. This reflects the limited specificity in currently used radiation detection instruments which are not sufficiently able to distinguish electron signatures associated with gamma- and beta-ray-emitting radionuclides and beta-only-emitting radionuclides. We introduce a new concept that allows the fast determination and discrimination of beta- and gamma-ray emitting radionuclides in an unknown sample. Of particular interest is the ability to detect and measure non-gamma ray emitting radionuclides such as Sr-90 with sufficient sensitivity and specificity to potentially a factor of ten below the Derived Response Level (DRL), which is important in response to radiological incidents \cite{epa2007}. In the following, we introduce this new concept, the instrument and the detection system, the modeling and simulations and their benchmarking with measurements, and the spectral decomposition, which allows the detection, quantification, and uncertainty estimation of radionuclides. We will then show and discuss measurement results with an unknown soil sample spiked with fission products.

\section{Approach}
\label{sec:app}

\begin{figure*}[!htb]
   \centering
   \includegraphics[width=1.0\textwidth]{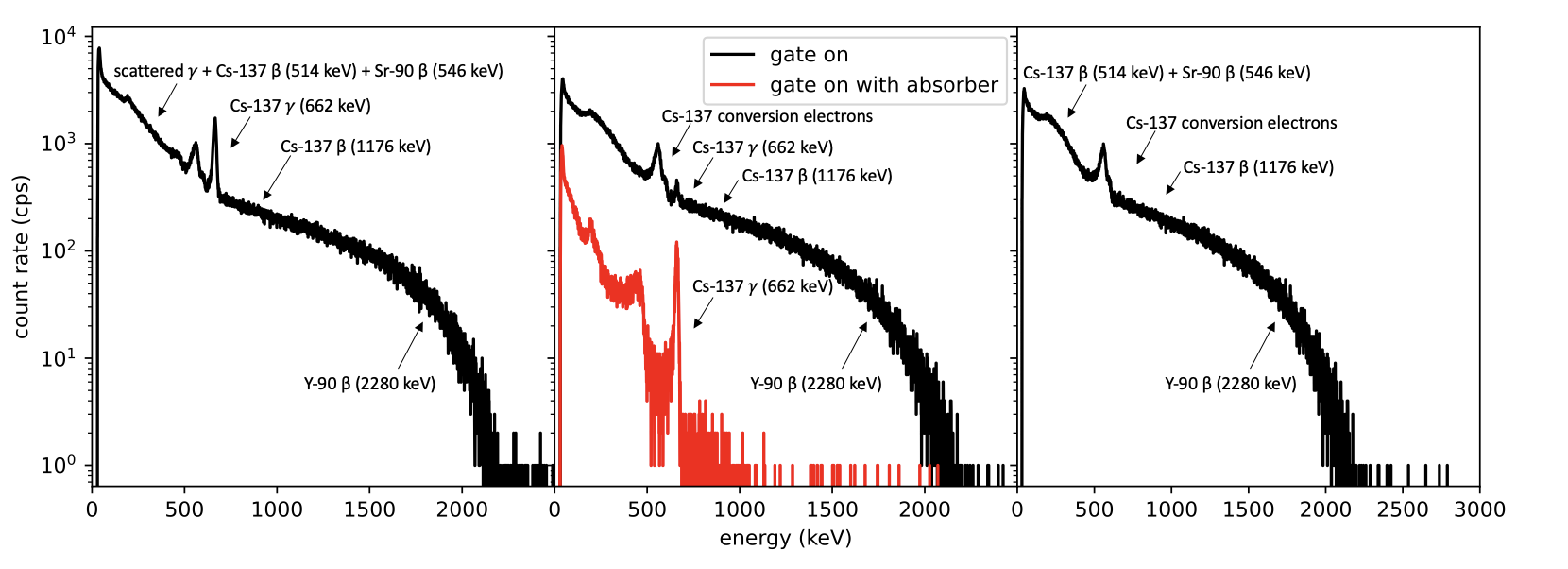}
   \caption{Spectra obtained with a commercial CPG CZT detector exposed to a mixed source of Cs-137 and Sr-90 are shown. On the left is the full spectrum; in the middle, the spectra gated on events in close proximity to the front, without (black) and with (red) a 3 mm Al absorber; on the right is the difference of the gated spectra resulting in a “clean” electron spectrum with the gamma-ray background from the sample and the environment removed.}
   \label{fig:concept}
\end{figure*}

Room-temperature operational semiconductor detectors such as CdZnTe (CZT) have undergone remarkable advances over the last 20 years in the improvement of crystal growth, detector fabrication, and signal processing. In particular, the realization of segmented electrode readouts enables improved efficiency, energy resolution, and depth-of-interaction (DOI) determination over non-segmented detectors \cite{luke1995,he1997,luke2000}. The coplanar-grid (CPG) implementation provides good energy resolution of about 1.5\% at 662 keV with only one simple readout channel. Using this readout channel and the readout of one of the two grid electrodes or the cathode, a simple implementation of the DOI is possible. The combination of commercially available CPG CZT detectors equipped with a thin entrance window provides excellent sensitivity and energy resolution in the simultaneous detection of gamma rays and electrons. In principle, the measurement of the full energy of the gamma rays in the detector allows for the estimation of the gamma-ray emitting radionuclides and, with that, the estimation of beta-decay induced and conversion electrons emitted from them and, consequently, the determination of the beta-only emitting spectral component. However, in order to improve the ability to separate electron-only emitting radionuclides, we utilize the DOI. We divide the detector into two volumes: one with a shallow depth, situated close to the radiation source (at the front) that experiences both gamma-ray and electron interactions, and another volume encompassing depths beyond the range of electrons, which exclusively registers gamma-ray interactions. In addition, we introduce an absorber between the sample and the detector with sufficient thickness to attenuate the electrons emitted from the sample. By conducting measurements both with and without the absorber, we obtain four distinct spectra: a DOI-gated spectrum that captures all interactions in the front, a spectrum without this DOI gate, and both with and without the absorber. We illustrate the concept in Fig.~\ref{fig:concept} for the measurements of a mixed sample of Cs-137 and Sr-90. Sr-90 decays into Y-90 with a half-life of 29 years and a maximum beta energy of 546 keV. 

However, Y-90 has a half-life of just 64 hours, so it is seen alongside Sr-90 coming to secular equilibrium in about two weeks. Y-90 decays with a maximum beta energy of 2280 keV. Neither of these isotopes emits any significant gamma rays. As observed after the Chernobyl accident in April of 1986, Sr-90 needs to be distinguished from Cs-137 as both are emitted in fission with large yields and both are characterized by half-lives of the order of 30 years. Cs-137 decays via beta decay with a maximum beta energy of 514 keV (94\% of decays). In this case, it also usually emits a distinctive gamma-ray of 662 keV; otherwise, it emits conversion electrons of about 630 keV. In the other 6\% of decays, it undergoes beta decay to a ground state, with a maximum beta energy of 1176 keV. As seen in Fig.~\ref{fig:concept}, the energy resolution and thin window of the CZT detector allow for high-resolution energy spectra of both beta and gamma radiation. The narrow peak at 662 keV is produced by gamma radiation from Cs-137, while the broader peak around 500 keV is from the Cs-137 conversion electrons. Many of the events below 500 keV are beta emissions from either Cs-137 or Sr-90. Above 662 keV, some events are from the higher-energy Cs-137 beta emission, but the distribution that extends up to 2.2 MeV consists mainly of Sr-90’s daughter Y-90 beta decays. Some gamma-ray-induced background is observed throughout the spectrum. A simple depth discrimination circuit vetoes any event that is not near the cathode of the CZT (the surface nearest the sample) and results in the spectra shown in the center (black) in Fig.~\ref{fig:concept}. In this way, we can enhance the electron signal by suppressing the gamma-ray signal, as most electrons from the sample are stopped within a short distance. The spectrum in red is obtained by the DOI gating and employing a 3 mm thick Al absorber between the sources and the detector. The Al absorber attenuates the electron signal, and the spectrum is dominated by the gamma-ray interaction in the thin detector layer close to the cathode. This spectrum reflects the background spectrum that can be subtracted from the gated spectrum obtained without the absorber, resulting in the spectrum shown on the right of Fig.~\ref{fig:concept}. This spectrum reflects a filtered electron spectrum where all the expected signatures associated with the beta decay and conversion electrons are visible without any significant gamma-ray background. It demonstrates the underlying concept of measuring electron signatures of samples.

While it is possible to remove the gamma-ray background and obtain a clean electron signal with this method, as shown in Fig.~\ref{fig:concept}, it is limited in the detection and accurate concentration estimation, particularly in more complex samples. Therefore, we developed a new methodology to decompose measured spectra into spectra that have been created with simulations for a list of individual relevant radionuclides and backgrounds. This approach allows us to increase the sensitivity and specificity in the extraction of beta- and gamma-ray emitting radionuclides, enables the quantification of radionuclides in samples, and enables the determination of detection limits.

\section{Instrument}
\label{sec:instr}

\begin{figure}[!htb]
   \centering
   \includegraphics[width=1.0\textwidth]{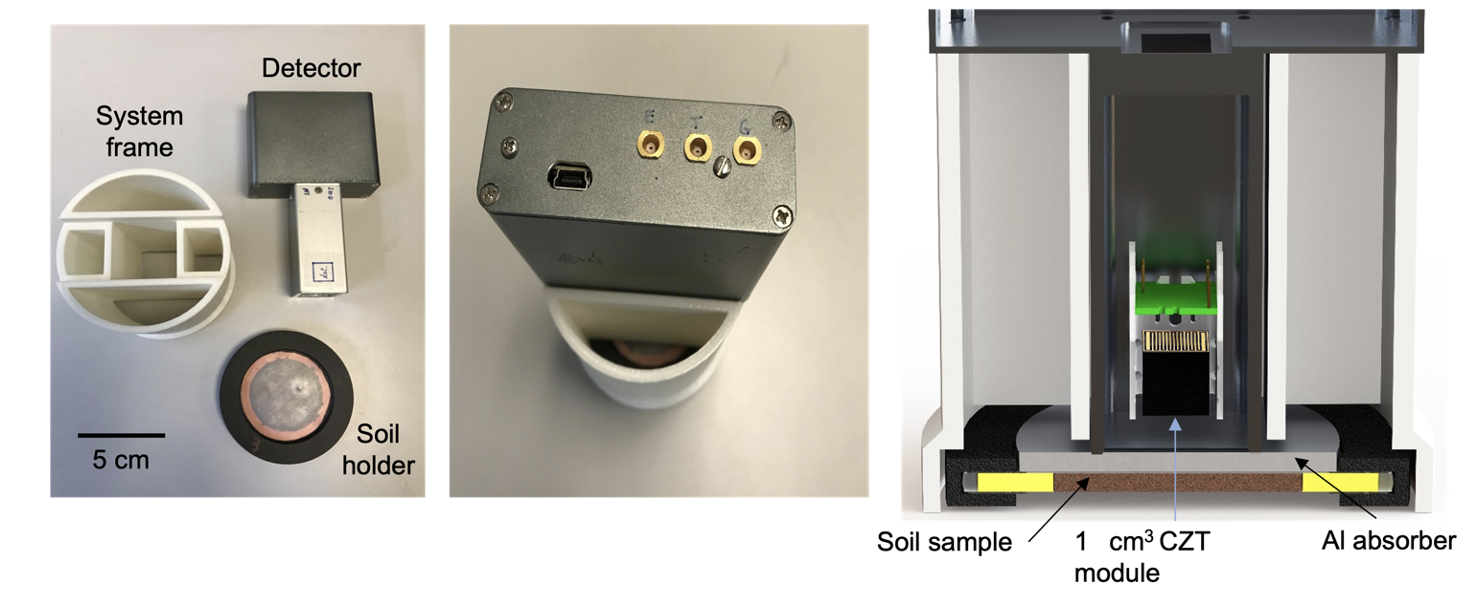}
   \caption{The detection system including detector, readout electronics, source holder, and mechanical support are laid out separately (left) and integrated together (middle). A view of the detection system model with the CZT crystal, housing, readout, mechanical support, and source holder with source is shown (right).}
   \label{fig:ins}
\end{figure}

We have adapted a CPG CZT detector commercially available as the GR-1 by KROMEK \cite{kromek-GR1}. We thinned the Al entrance window of the detector housing to 16~$\mu$m and added a DOI circuit that relates the corrected signal amplitude to the collecting-grid signal amplitude. Fig.~\ref{fig:ins} shows the components of our simple setup, including the detector, the sample holder, and the mount which holds both the detector and the sample holder. The sample holder is comprised of two plastic disks, each surrounding an attached thin mylar foil. The disks can be screwed together, leaving a distance of 3 mm between the two foils, defining the dimensions of the sample to be 3 mm in thickness and 48 mm in diameter. Various types of soil, mixed materials, or filters can easily be loaded into this holder and mounted in the system. The dimensions of the sample holder were chosen to optimize the trade-off between maximizing the amount of material and minimizing the attenuation and energy loss of emitted electrons. The geometry of the overall setup, including the area of the detector, was taken into account. While we present results obtained from a single GR-1 type detector which is (1x1x1) cm$^3$ in size, in future implementations, we envision using at least one larger-area detector on each side of the sample, which should result in a two- to ten-fold increase in sensitivity, depending on the number and sizes of the detectors. There is space between the sample holder and the detector for a removable absorber used for the measurements described below and sufficient to attenuate the electrons emitted in the beta decay of Y-90 with its endpoint energy of 2279 keV. The detection system has overall dimensions of about (5x5x12) cm$^3$ and is powered, controlled, and read out via a USB interface.

\section{Modeling and Simulations}
\label{sec:modeling}

In order to determine the isotopic composition of the unknown sample, we decompose each of the measured energy spectra introduced above into simulated basis spectra of individual radionuclides. These basis spectra are derived by simulating for each radioisotope its emission of gamma rays and/or beta particles and their interaction in the CZT detector, providing a list of energy depositions and positions. This information is then combined with a calculated and calibrated detector response which takes into account charge transport, charge loss, electronic response, and noise in the CPG CZT detector. This ultimately results in the calculated energy spectra \cite{galloway2011}. The MEGAlib framework has been used to perform the Monte-Carlo simulations and include the detector effects \cite{zoglauer2006}. 
The detector responses were obtained experimentally, with scans of a Cs-137 source along the depth of the detector using a collimator with an opening of 0.5 mm.

\begin{figure*}[!htb]
   \centering
   \includegraphics[width=1.0\textwidth]{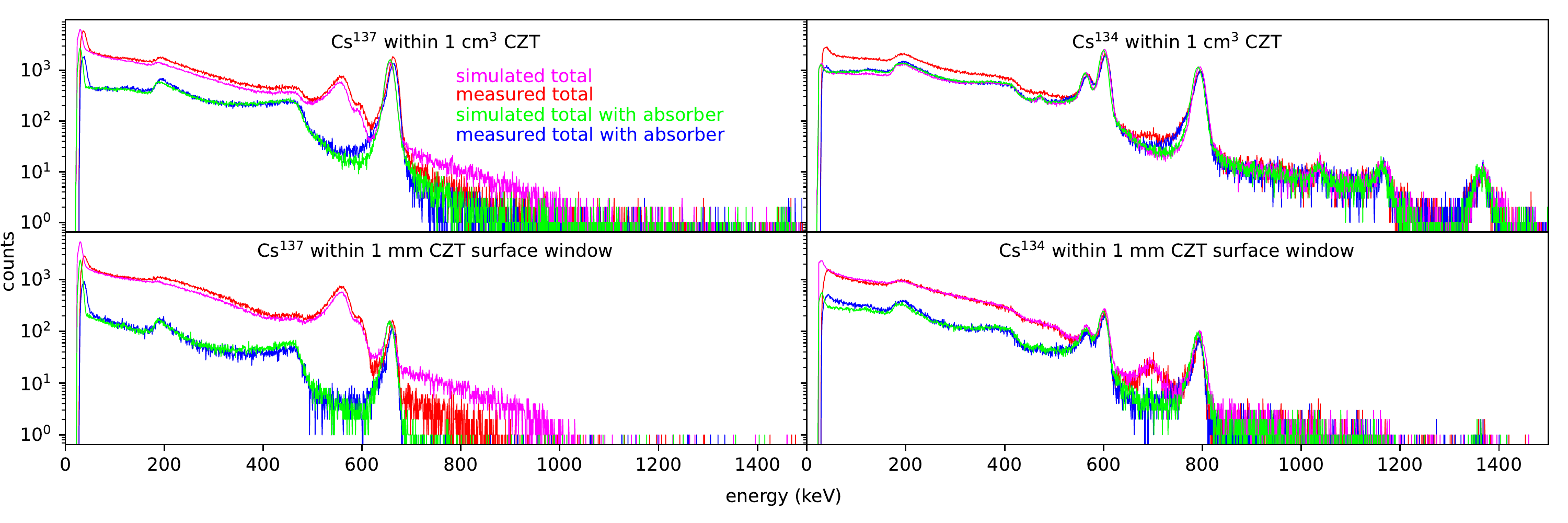}
   \caption{Comparisons of measured and simulated energy spectra obtained in response to Cs-137 (left) and Cs-134 (right), as a result of the emitted gamma rays, conversion electrons, and electrons emitted in the beta-decays, are shown. The spectra are created assuming the same number of disintegrations from the source. The top plots show spectra obtained within the volume of the 1 cm$\times$1 cm$\times$1 cm CdZnTe detector; The bottom plots show spectra obtained for the 1 mm thick front layer of the same CdZnTe detector.}
   \label{fig:cs-comp}
\end{figure*}

\begin{figure*}[!htb]
   \centering
   \includegraphics[width=1.0\textwidth]{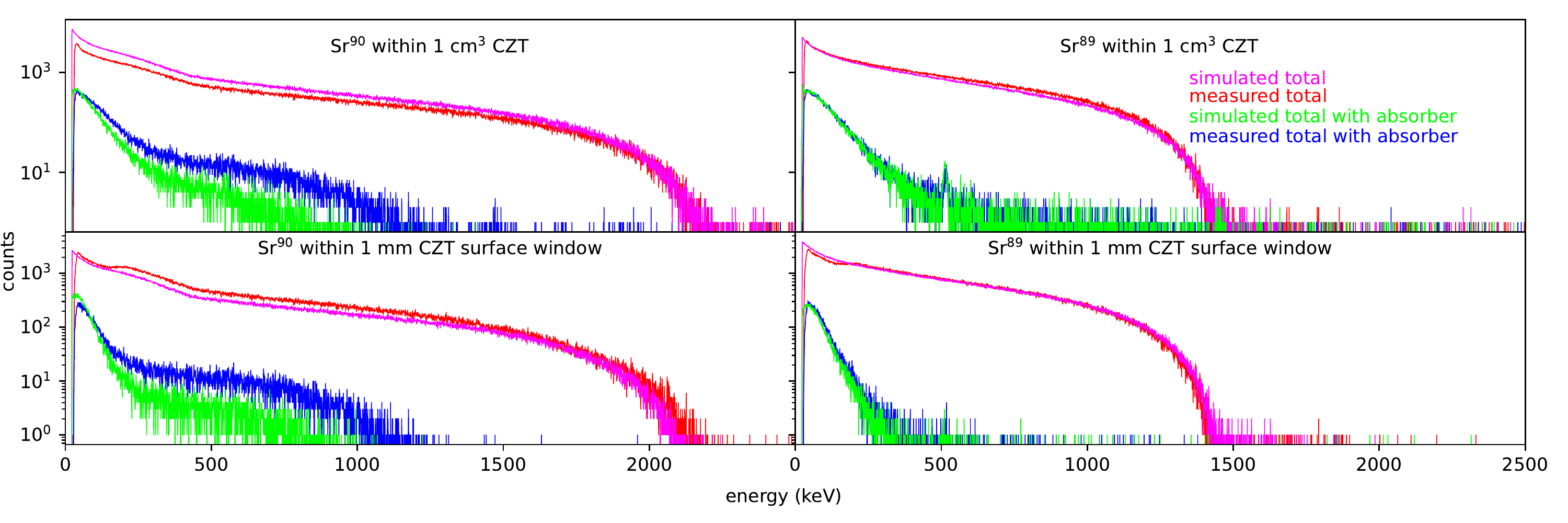}
   \caption{Comparisons of measured and simulated energy spectra obtained in response to Sr-90(left) and Sr-89 (right) are plotted. The spectra are created assuming the same number of disintegrations from the source. The top plots show spectra obtained within the volume of the 1 cm$\times$1 cm$\times$1 cm CdZnTe detector; The bottom plots show spectra obtained for the 1 mm thick front layer of the same CdZnTe detector.}
   \label{fig:sr-comp}
\end{figure*}

Figs. \ref{fig:cs-comp} and \ref{fig:sr-comp} show examples of comparisons between simulated and measured spectra for Cs-134 and Cs-137 and Sr-89 and Sr-90, respectively. The measurements and simulations were performed with two different types of sources provided by Eckert\&Ziegler. The first type consisted of beta particle standards with thin aluminized mylar windows, an active diameter of 20.4 mm, and an activity of 100 nCi. The second type of sources - made only with Sr-90 - consisted of two custom-made soil matrices\footnote{10 mg of pulverized Griffin soil or 15 mg of sieved Macon soil, each fitting in the 5.4 ml of the sample holder.} spiked with activities of 1 nCi or 50 nCi. Overall, a good agreement between the measurements and the simulations can be observed, providing confidence in the decomposition of measured with the simulated basis spectra. Both clearly show the gamma-ray-induced Compton-scattering background and full-energy peaks, as well as the conversion electrons in Cs-134 and Cs-137 and the beta spectra for all present radionuclides. Also visible is the desired suppression of the electrons due to the absorber and the fact that most of the electron energies are deposited in the front layer of the CZT detector resulting in a relative enhancement of the electron spectral component as compared to the gamma-ray component, which motivates the use of the four specific spectra. 

In order to further improve the sensitivity to signatures related to electrons, we take the attenuation and energy loss of electrons in the sample into account by dividing the source into depth slices in the simulations. Furthermore, since we envision ultimately using a system consisting of two detectors with the sample in the middle, we utilize the simulated spectra twice. Eventually, using two opposing detectors will allow us to double the sensitivity and minimize systematic uncertainties related to the varying and energy-dependent ranges and specific energy losses of electrons in the sample in the spectral decomposition. 

\section{Spectral Decomposition}
\label{sec:specdecom}

In order to estimate the concentrations of a wide range of gamma-ray and beta-emitting radionuclides, a spectral decomposition methodology was developed that allows the determination of the concentrations and their uncertainties as well as detection limits for currently up to 25 different radionuclides. The detection system discussed above measures 8 spectra of samples over an energy range of 51-3100 keV represented in 4031 channels and two depth regions, one near the surface and one throughout the bulk of the two detectors: (1) spectra from both sides of the sample, (2) with and without absorber, and (3) with and without depth discrimination. The depth discrimination within the detector is used to distinguish between the gamma radiation that is absorbed fairly uniformly throughout the detector and the beta radiation that is preferentially absorbed near the surface. 

In principle, the spectral decomposition algorithm should be able to use these spectra to determine the concentrations of the various radionuclides within the sample. However, this simple task is complicated by the sensitivity of the beta spectra to attenuation and straggling, so the observed spectra can vary significantly as a function of the source depth within the sample. Because larger sample sizes are required for good counting statistics, the sources are necessarily thick enough for the depth effects to alter the observed spectra. Thus, the algorithm must account not only for the concentration of the radionuclides in the sample but also for the depth dependence of those isotopes for the particular sample thickness and sample density. For this reason, the sample thickness and density are measured, and the radionuclide concentrations must be reconstructed as a function of depth within the sample. Therefore, the algorithm for the determination of 25 radionuclides requires the reconstruction of the source concentrations as a function of depth. In our approach, we define five different depth ranges, which implies the determination of 125 source concentrations instead of 25. Fortunately, the 8 spectra provide 32248 channels of information - which seems potentially enough spectral data for the complete reconstruction of 125 source concentrations.  

Because the spectral data are collected over a fixed time period, the counts in any energy channel are Poisson distributed. The mean number of counts in each spectral channel is a linear combination of the radioisotope concentrations at the various depths. Therefore, we can build a system matrix that describes the mean counts observed in the spectra in terms of the isotopic concentrations. This systems matrix contains ($32248 \times 125$) elements, M$_{\alpha i}$, that describe the contribution of each isotope at a particular depth ($i=1,…,125$) to the spectral element ($\alpha=1, ..., 32248$). For such a Poisson process with a known system matrix, the Maximum Likelihood Expectation Maximization (MLEM) methodology is well suited for the determination of the sources. 

In actual measurements, counts from background radiation are indistinguishable from those of the source sample. To account for this background, we include additional “background” sources for each detector. Since background measurements are susceptible to systematic uncertainties, specifically when taken in the field, two default background spectra from different locations and environments are taken into account as well. The current algorithm is designed for the inclusion of either in-situ or default backgrounds. For this purpose, three background sources are included so that, rather than 125 sources, there are 128 source channels. The algorithm decides the appropriate combination of background models that provides the best fit for the data. When the MLEM algorithm is applied, the background generally contributes only a small part to the observed spectra. As a result, small fluctuations in the measured spectra can produce large changes in the assigned background. To suppress fluctuations in the reconstructed background, we introduced penalty functions designed to assure the total backgrounds were within the range of previous measurements. Two additional measurement channels were added to assure that these constraints were governed by the Poisson statistics of the data. With the addition of these channels, the system matrix grows to ($32250 \times 128$).

The MLEM algorithm was applied to numerous experimental measurements. The initial implementation did not give realistic or reliable results. The reason for this failure was immediately recognized: while there were abundant channels of (noisy) data, many of the sources produced nearly identical signals. The MLEM algorithm amplified signals that produced small changes in spectra in order to fit the details of the “noisy” data. This diagnosis was verified with a Singular Value Decomposition (SVD) analysis of the system matrix – which revealed that 20-30 channels of information were recoverable from the noisy data, whereas the MLEM was attempting to recover 128. There seemed to be sufficient information to recover 25 isotopes but not enough to handle the source depth simultaneously. Essentially, the MLEM was amplifying low-sensitivity sources to fit noise. The problem was resolved by the introduction of a penalty function that suppressed low-sensitivity sources. This penalty function can be viewed as a regularization term that inhibits the amplification of noise. We anticipated the introduction of additional penalty functions, but suppression of low-sensitivity sources was sufficient for the prototype device to give reasonable results. Nonetheless, additional or alternate penalty functions are available, for example, to suppress depth variations of source concentrations or to amplify known isotopic ratios as prior information. 

Once the MLEM algorithm with the penalty function regularization was implemented, the results seemed realistic and accurate when compared with known test samples. However, because the MLEM does not immediately provide uncertainties for the reconstructed source concentrations, quantitative comparisons were problematic. Fortunately, some straightforward analysis yields the errors associated with the reconstructed sources. Basically, small Poisson variations in the observed spectra cause shifts in the Likelihood function that, in turn, require shifts in the reconstructed sources so that the sources continue to maximize the Likelihood. We find a linear relation between the variations in the observed spectra and the shifts in the reconstructed sources. This relation allows us to calculate the covariance matrix between the various source concentrations caused by the (independent) Poisson variations observed in the spectra. The standard deviations of the sources are given by the diagonal entries of this covariance matrix. Computationally, these error estimates require the SVD inversion of the system matrix – which is easy for a relatively small ($32250 \times 128$) system matrix. This method of error estimation was tested and verified using Poisson simulations with the system matrix. When applied to the reconstruction of known sources, the error analysis gave results consistent with the observed sources. Furthermore, these error estimates permit the determination of limits on detectable source concentrations, which vary from isotope to isotope.

\section{Results}
\label{sec:results}

In the following, we present results obtained in the analysis of a third type of source provided by Eckert\&Ziegler. Eckert\&Ziegler prepared two samples of soil spiked with fission fragments. After mixing the samples, two holders were filled with soil from both samples and subsequently measured on the day of arrival. A second set of measurements were performed about three weeks later to study the impact of the physical decays of some of the fission fragments.

\begin{figure}[htp]
\centering
\subfloat{
\includegraphics[clip,width=0.8\columnwidth]{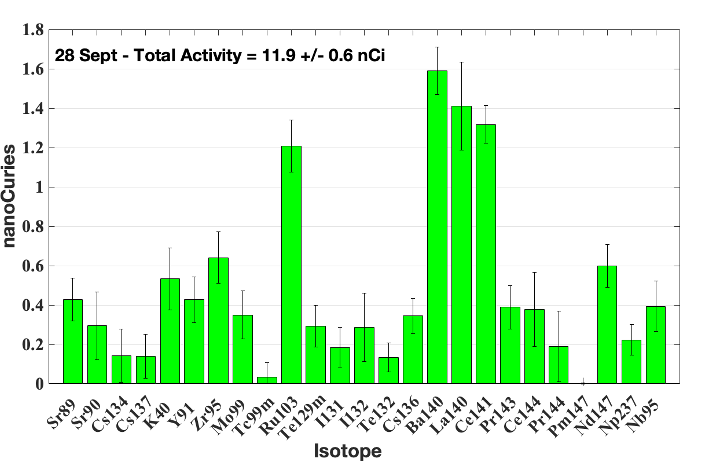}
} \\
\subfloat{
\includegraphics[clip,width=0.8\columnwidth]{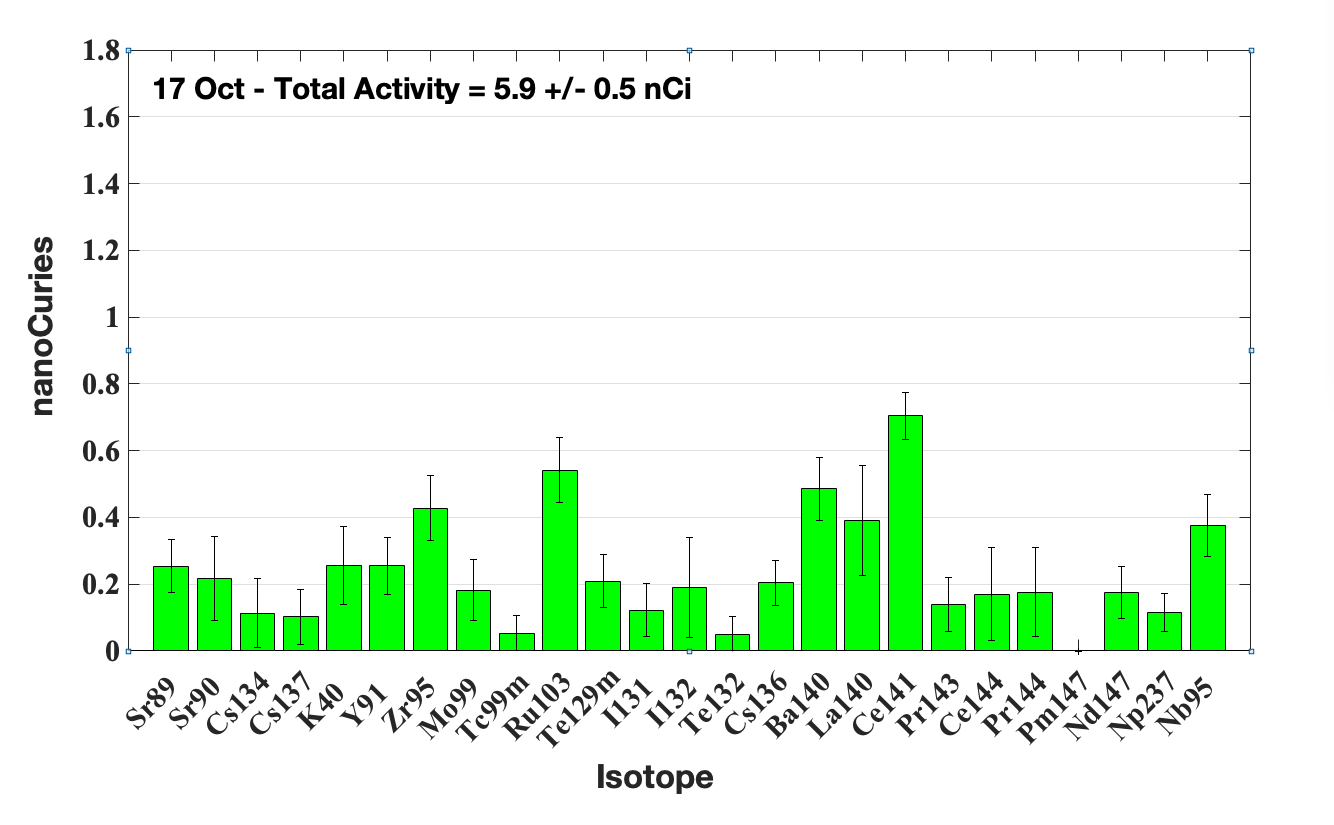}
}

\caption{Reconstructed activities of radioisotopes in an unknown soil sample that was spiked with fission fragments provided by Eckert\&Ziegler are shown. The figures show reconstructions from measurements that were taken 19 days apart.}
\label{fig:speccomp}

\end{figure}

Fig.~\ref{fig:speccomp} shows the results of the reconstructed radioisotope-specific activities with their uncertainties of one of the samples measured 19 days apart. It is interesting to note the changes in the activities of the strongest identified contributions of Ba-140, La-140, and Ce-141 track the half-lives of 12.7 d, 1.7 d, and 32.5 d, respectively.  

\begin{figure}[!htb]
   \centering
   \includegraphics[width=1.0\textwidth]{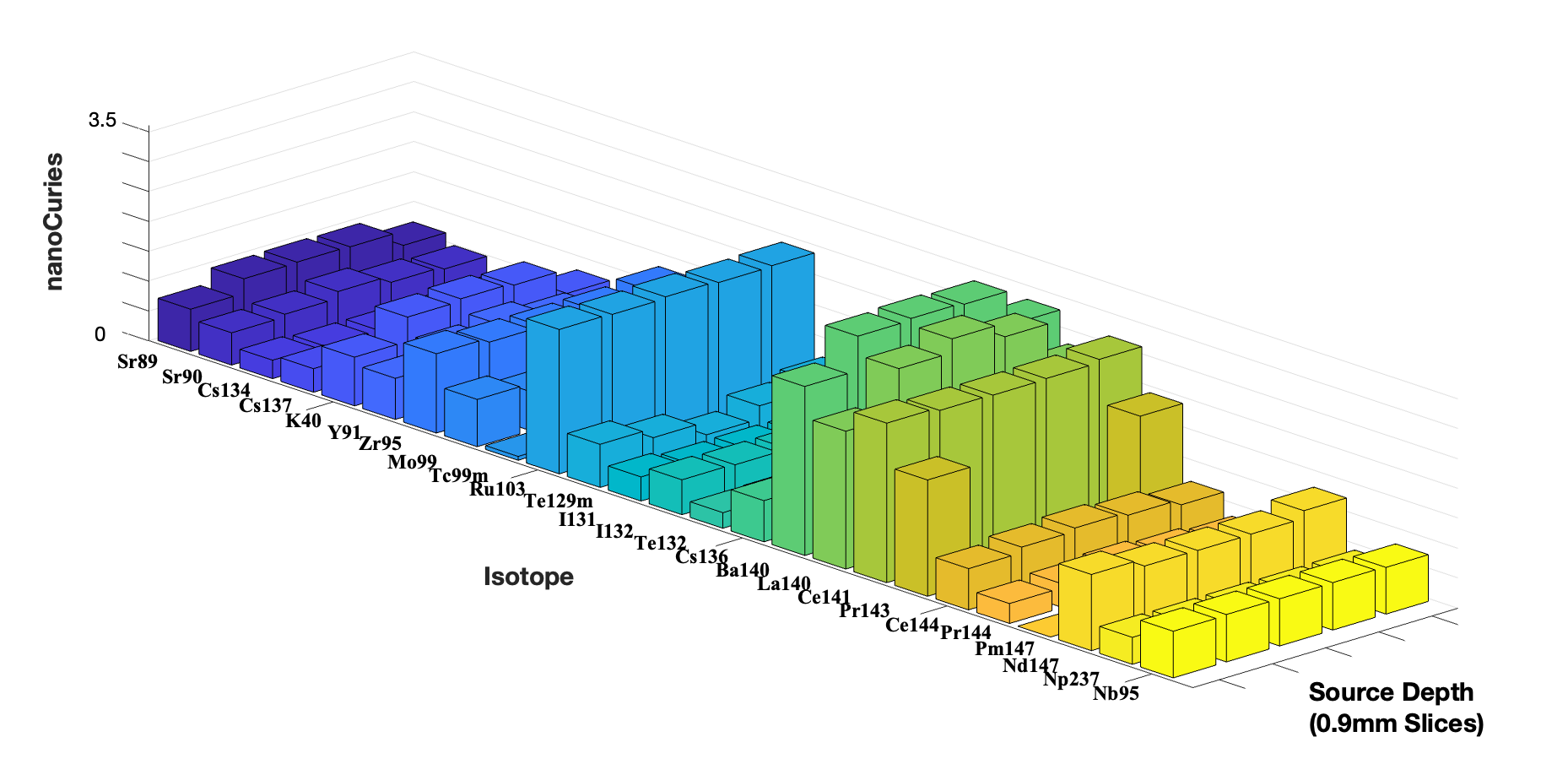}
   \caption{Activities were reconstructed for 5 depth bins of the unknown soil sample for each radioisotope.}
   \label{fig:depth}
\end{figure}

Fig.~\ref{fig:depth} shows the results of the analysis of one sample in more detail and according to the reconstruction of the depth. This figure provides an additional degree of freedom (source depth) that reveals a potentially problematic reconstruction. For example, the reconstruction of Pr-143 is biased towards the surface of the sample, which is not physical and, therefore, might point to a systematic error in the reconstruction for this isotope. 

\begin{figure*}[!htb]
   \centering
   \includegraphics[width=1.0\textwidth]{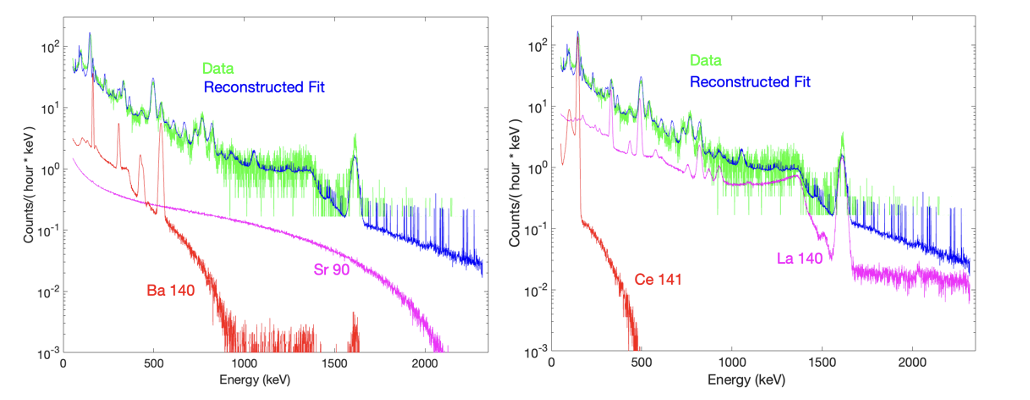}
   \caption{Reconstructed energy spectra are shown for the three strongest activities and Sr-90. The green and blue spectra represent the measurements and the fit without the absorber, respectively. The left figure also shows the reconstructed spectral components of Ba-140 and Sr-90 in red and magenta, respectively. The figure on the right shows the reconstructed spectral components of Ce-141 and La-140 in red and magenta, respectively.}
   \label{fig:Ba140recon}
\end{figure*}

Fig.~\ref{fig:Ba140recon} shows reconstructed energy spectra in comparison with the measured spectra. Specifically, the strongest radionuclides (Ba-140, La-140, and Ce-141) are shown with their deposited energies from the gamma rays and electrons (from beta decay and conversion) as well as the Sr-90 spectrum due to the electrons from the beta decays, only. 

\begin{figure*}[!htb]
   \centering
   \includegraphics[width=1.0\textwidth]{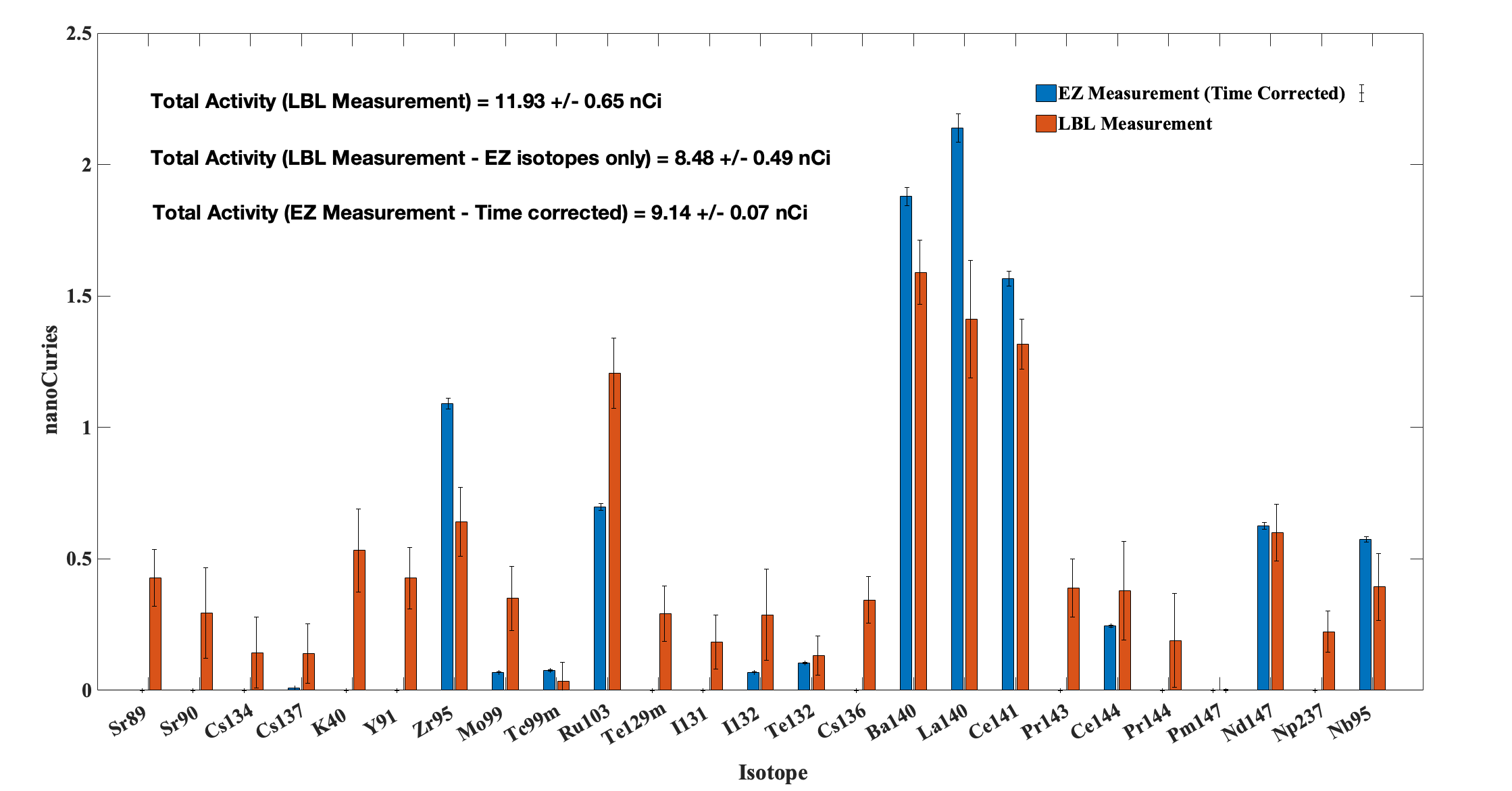}
   \caption{Reconstructed activities of all identified radionuclides are compared with values provided by the manufacturer. In addition to the radionuclides provided by the manufacturer Eckert\&Ziegler, we are able to observe several additional radioisotopes and the non-gamma-ray emitting Sr-89 and Sr-90.}
   \label{fig:recon}
\end{figure*}

\begin{figure*}[!htb]
   \centering
   \includegraphics[width=0.5\textwidth]{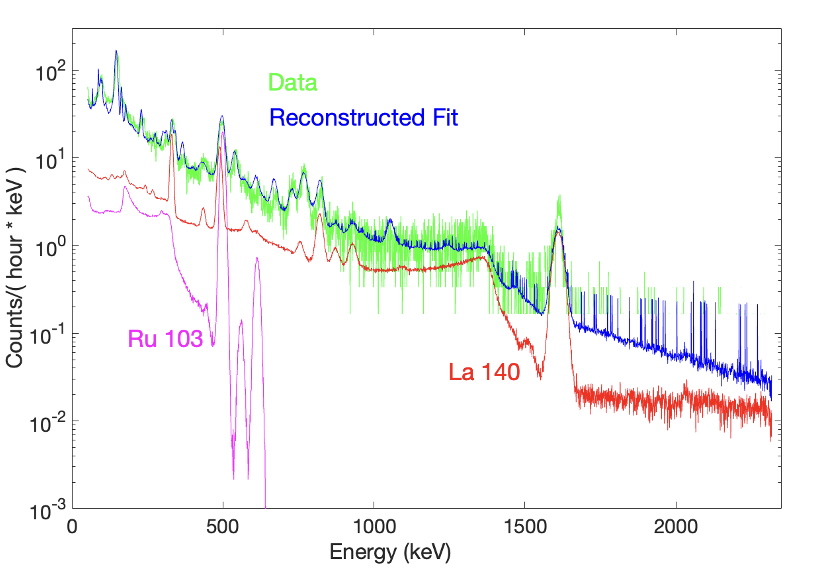}
   \caption{The total energy spectrum measurement is shown in green, and the reconstructed spectrum is shown in blue. The La-140 and Ru-103 components of the reconstructed fit are shown in red and magenta, respectively. A comparison of these components in the vicinity of 490 keV reveals the challenge of distinguishing isotopes with overlapping emission lines.}
   \label{fig:Te132recon}
\end{figure*}

Reconstructed activities for all identified radionuclides with the recently developed methodology to estimate uncertainties are shown in Fig.~\ref{fig:recon} compared to the values provided by Eckert\&Ziegler. In addition to determining the activity of the gamma-ray emitting radionuclides, we can also determine the beta-only-emitting radionuclides such as Sr-90 and Sr-89. The manufacturer only measured and provided activities for gamma-ray-emitting radionuclides. Overall, we find good agreement, except for cases such as Pr-143 and Ru-103. In the case of Pr-143, the reconstructed depth distribution seen in Fig.~\ref{fig:depth} indicates that there is a systematic error in the reconstruction of this radioisotope. In the case of Ru-103, there is an overlap in the spectral features of La-140 and Ru-103, as seen in Fig.~\ref{fig:Te132recon}, which complicates reconstruction and causes understood errors. One might think that the photopeak of La-140 at 1596 keV could be used to differentiate it from Ru-103. However, this photopeak is more than an order of magnitude smaller than the peaks at 487 keV (La-140) and 497 keV (Ru-103), which overlap in the total spectrum. The Poisson statistics implicit in MLEM favor the overlapping photopeaks. When faced with such a situation, MLEM balances the distribution between the two possible solutions equally; thereby, overestimating Ru-103 and underestimating La-140, as seen in Fig.~\ref{fig:recon}.

The manufacturer declared the gamma-emitter activity (corrected for decay time) to be 9.14$\pm$0.07 nCi for the 13 g sample. We calculate the total activity to be 11.93$\pm$0.65 nCi. This total reconstructed activity is higher than the activity provided by the manufacturer. However, when we only take into account the radionuclides measured by the manufacturer, we obtain a reconstructed activity of 8.48$\pm$0.49~nCi which shows good agreement. As examples, the reconstructed activities of Sr-90 and Ba-140 correspond to 7$\pm$3~kBq/m$^2$ and 32$\pm$2~kBq/m$^2$, respectively. 

\section{Conclusion}
\label{sec:conc}

The rapid, in-field assessment of beta-only emitting radionuclides, such as Sr-90, remains an outstanding challenge, particularly in radiological decontamination, decommissioning, emergency response, and consequence management situations. Current approaches for the quantification of pure beta-emitting isotopes involve sending the sample to a lab for chemical separation and measurements, which can take several weeks. We present a new approach consisting of a simple and hand-portable instrument capable of measuring beta-only emitting radionuclides in environmental samples such as soil or debris quickly and in-situ or in a Fly-Away Lab as used by the U.S. Nuclear Incidence Response Teams. The instrument consists of CZT detectors in the CPG configuration, with minor modifications and new signal readouts, in a fixed geometry. 

With a single detector prototype, the underlying concept was successfully demonstrated and provided activities of all fission products within the given uncertainties and in good agreement overall with the activities provided by the vendor, Eckert\&Ziegler. Within 30 minutes, Sr-90 was measured at a level of 300+/-100~pCi in the 13~g soil sample. Currently, a fully operational instrument is being built with two larger-area detectors and anticipated to have increased sensitivity by a factor of 10. This can be translated to a sensitivity down to 150~pCi/kg or less than 6 Bq/kg in 30 minutes. Increasing the total detector surface by adding more detectors is possible, to gain another factor of 10 increase in sensitivity. 

With regard to surface contamination, we achieve a sensitivity for Sr-90 of 4$\times$10$^5$~Bq/m$^2$ in less than 60~minutes with the current one-detector system and a detector area of 1~cm$^2$. With a fully operational two-detector system, each detector with an area of 4~cm$^2$, we expect to achieve a sensitivity of 3.7x10$^4$Bq/m$^2$ in less than 60~minutes which is 10\% of the Derived Response Level or DRL and of importance to inform the emergency response actions \cite{frmac2019}.  

While this specific system was designed for in-field assessment of beta-only emitting radionuclides for consequence management, the flexible design should be able to accommodate different requirements, e.g. relevant for decontamination and decommissioning activities. This instrument could be used as an effective screening instrument for a wide range of samples for in-situ assessment, preventing the lengthy, albeit more precise and more sensitive, remote assessment. In addition, when combined with the recently developed Scene-Data Fusion concept \cite{vetter2019}, a remote deployment on an unmanned system where the instrument can be brought close to potentially contaminated surfaces in indoor and outdoor environments appears feasible. 

\section{Acknowledgements}
\label{sec:ack}
This work was performed under the auspices of the U.S. Department of Energy by Lawrence Berkeley National Laboratory under Contract DE-AC02-05CH11231. The project was funded by the U.S. Department of Energy/National Nuclear Security Administration’s (DOE/NNSA’s) Office of Nuclear Incident Response Program (NA-84).



\bibliographystyle{elsarticle-num} 
\bibliography{Fermi-Beta}


\end{document}